\documentclass{article}

\usepackage{amssymb}
\usepackage{amsthm}
\usepackage{amsmath}

\usepackage{graphicx}

\newtheorem{theorem}{Theorem}[section]
\newtheorem{prop}[theorem]{Proposition}

\newtheorem{conj}[theorem]{Conjecture}
\newtheorem{remark}[theorem]{Remark}

\newcommand{\obeta}{{\overline{\beta}}}
\newcommand{\wW}{\widetilde{W}}

\begin{document}
\title{Quasi-classical approximation in vortex filament dynamics. Integrable systems, gradient catastrophe and flutter}
\author{
B.G.Konopelchenko $^1$ and G.Ortenzi $^2$\\
$^1$ {\footnotesize Dipartimento di Matematica e Fisica {``Ennio De Giorgi''}, Universit\`{a} del Salento }  \\
{\footnotesize and INFN, Sezione di Lecce, 73100 Lecce, Italy, \texttt{konopel@le.infn.it }} \\
 $^2$ {\footnotesize Dipartimento di Matematica Pura ed Applicazioni, } \\
{\footnotesize Universit\`{a} di Milano Bicocca, 20125 Milano, Italy, \texttt{giovanni.ortenzi@unimib.it} }
}
\maketitle
\begin{abstract}
Quasiclassical approximation in the intrinsic description of the vortex filament dynamics is discussed. Within this approximation the governing equations are given by elliptic system of quasi-linear PDEs of the first order. Dispersionless Da Rios system and dispersionless Hirota equation are among them. They describe motion of vortex filament with slow varying curvature and torsion without or with axial flow. Gradient catastrophe for governing equations is studied. It is shown that geometrically this catastrophe manifests as a fast oscillation of a filament curve around the rectifying plane which resembles the flutter of airfoils. Analytically it is the elliptic umbilic singularity in the terminology of the catastrophe theory. It is demonstrated that its double scaling regularization is governed by the Painlev\'e I  equation. 
\end{abstract}
\tableofcontents
\section{Introduction}
Dynamics of vortices has been the subject of intensive study since the pioneering works of Helmholtz, Lord Kelvin and Poincar\'e in the 19th century (see e.g. \cite{Bat,Saf}). Relatively simple dynamics of a thin vortex filament in an incompressible inviscid fluid and in an infinite three-dimensional domain has attracted a particular interest. \par
An approach based on the localized induction approximation (LIA) has been proposed in 1906 by Da Rios \cite{DR}. Within such approximation one assumes that the vorticity $\vec{\omega}$ is confined to a thin vortex core region and the core radius $a$ of the vortex tube is much smaller than the radius of curvature $R$ at the point and, hence, the induced velocity at this point is dominated by the contribution from the neighboring segment. Using the Biot-Savart formula
\begin{equation}
\label{BS-law}
 \vec{v}= -\frac{1}{4 \pi} \int \vec{\omega}(\vec{x'}) \times  \frac{\vec{x}-\vec{x'}}{|\vec{x}-\vec{x'}|^3}\ d\vec{x'},
\end{equation}
 under the assumption that the circulation $\Gamma$ of the vortex is a constant, one gets
\begin{equation}
\label{bin-vel}
 \vec{v}(s,t)= \frac{\partial \vec{X}(s,t)}{\partial t}=\frac{\Gamma }{4 \pi} \ln\left(\frac{L}{a}\right)K \vec{b}  
\end{equation}
where $\vec{X}$ denotes a point on the filament, $s$ is the arclenght, $t$ is time, $K$ is curvature, $\vec{b}$ denotes the binormal vector and $L$ is the length of the segment of the filament whose contribution into the induced velocity is taken into account in the integral 
(\ref{BS-law}).  Assuming that one can neglect the variation of $L$ along the filament, rescaling time by the constant $\frac{\Gamma }{4 \pi} \ln\left(\frac{L}{a}\right)$ and using the expression of $\vec{b}$ in terms of $\vec{X}$, one obtains 
\begin{equation}
\label{bin-mot}
 \frac{\partial \vec{X}}{\partial t}=K \vec{b}=\vec{X}_s \times \vec{X}_{ss}. 
\end{equation}
 Thus, within LIA approach, one has a simple binormal motion of the filament with the induced velocity proportional to the curvature $K$. Using the Serret-Frenet formulae for the space curves, one can transform  (\ref{bin-mot}) into the system of intrinsic equations for the curvature $K$ and torsion $\tau$ which is of the form \cite{DR}
\begin{equation}
\label{bin-intrinsic}
\begin{split}
K_t&=-2K_s \tau -K \tau_s,\\
\tau_t &= KK_s -2\tau\tau_s + \left( \frac{K_{ss}}{K}\right)_s.
\end{split}
\end{equation}
These LIA and Da Rios system have been rediscovered sixty years later in \cite{Hama,AH,Bet}. The detailed historical account of this story is presented in \cite{Ric}.\par 
An important step in the development of the LIA approach has been done by Hasimoto in 1972 \cite{Has}. He observed that the system (\ref{bin-intrinsic}) is equivalent to the focusing nonlinear Schroedinger equation (NLS) 
\begin{equation}
 \label{NLS}
i \psi_t +\psi_{ss}+\frac{1}{2}|\psi|^2\psi=0
\end{equation}
via the transformation
\begin{equation}
\label{Has-trans}
 \psi(s,t)=K(s,t) e^{i \int ds' \tau(s',t)}.
\end{equation}
The NLS equation was already widely known in plasma physics. Moreover in 1971 Zakharov and Shabat \cite{ZS} discovered that NLS 
equation (\ref{NLS}) is integrable by the inverse scattering transform (IST) method. Hasimoto's result has demonstrated that the whole 
powerful machinery of the IST method (solitons, infinite sets of integrals of motion, symmetries etc.) is applicable to the vortex filament 
dynamics. This fact has lead to the explosion of interest to this field and since that time various aspects of the LIA in the vortex filament 
dynamics have been studied by different methods (see e.g. \cite{Lamb}-\cite{SC}). Instabilities of vortex filament and formation of singularities were two important problems partially addressed during this period \cite{KM,GRV,BV}.\par
At the same time it was clear that LIA is too restrictive since it neglects several important effects like axial flow, stretching, shortening, filament core structure etc. Attempts to go beyond LIA have been done in \cite{MS}-\cite{TN}. In particular, the inclusion of the axial flow into consideration has lead to the following expression for the induced velocity
\begin{equation}
 \label{eqn-axial}
\vec{X}_t = \alpha_1 K^2 \vec{t}+\frac{\Gamma}{4\pi} \left[ \ln\left( \frac{2L}{a}\right)+\alpha_2\right] K \vec{b} +\alpha_3 \vec{t} \times \vec{A} 
+ \alpha_4 \vec{t} \times (K\vec{b})_s+\vec{Q}
\end{equation}
where $\vec{t}$ is the tangent vector, $\alpha_i$ are some parameters defined by the called matching condition, $\vec{A}$ is a certain vector and $\vec{Q}$ denotes the contribution from the nonlocal effects. In situation when one can neglect $\vec{Q}$ and variation of $\frac{L}{a}$ along the filament, the vector $\vec{A}$ can be chosen as  $(K\vec{b})_s$. After a rescaling of time equation (\ref{eqn-axial}) takes the form \cite{FM}
\begin{equation}
\label{Eqn-axial-approx}
 \begin{split}
   \frac{\partial \vec{X}}{\partial t}&=K \vec{b} + \gamma_1 \left( \frac{1}{2}K^2 \vec{t} +K_s \vec{n} +K \tau \vec{b}\right) \\
&=\vec{X}_s \times \vec{X}_{ss} + \gamma_1 \left( \vec{X}_{sss} +\frac{3}{2} \vec{X}_{ss} \times  \left( \vec{X}_{s} \times \vec{X}_{ss} \right)\right)
 \end{split}
\end{equation}
where $\vec{n}$ is the normal vector, $\gamma_1$ is a certain parameter related to the intensity of the axial flow. After the Hasimoto transformation (\ref{Has-trans}) equation (\ref{Eqn-axial-approx}) becomes 
\begin{equation}
 \label{Hirota}
i \psi_t +\psi_{ss}+\frac{1}{2}|\psi|^2\psi +i \gamma_1 \left( \psi_{sss}+ \frac{3}{2}|\psi|^2\psi_s \right)=0
\end{equation}
that is the Hirota equation or the linear combination of the NLS and first higher NLS equation \cite{Hirota}. It is again an integrable equation.
On the other hand it describes new effects such as axial flow, local stretching proportional to $K_s$, binormal velocity depending on torsion $\tau$, effects which were not caught by LIA equation (\ref{bin-mot}). It was shown in \cite{FM} that axial flow affects essentially stability properties of the filament.The observation that the simple generalizations of equation (\ref{bin-mot})  or Da Rios system 
(\ref{bin-intrinsic}) may describe important nonlinear effects essential for vortex filament dynamics was one of the main results of \cite{FM}.
Another observation made in \cite{FM} based on the paper \cite{Bet} is that the study of the quasiclassical  (or dispersionless) limit of the equation (\ref{Hirota}) clarifies essentially an analysis of destabilizing effects in the vortex filament motion. \par
The latter result was strongly supported by the recent study of the modulation instability of the focusing NLS and higher NLS equations (see e.g. \cite{FL}-\cite{JMcL}). In these papers it was shown that the dispersionless limit of the NLS equation (dNLS) is quite relevant to the study of the modulation instability. In this limit NLS is the elliptic system of quasilinear PDEs which exhibits the gradient catastrophe behavior at finite time. Behavior of solutions near to the points of catastrophe and beyond has been studied in details in \cite{DGK}-\cite{JMcL}.  It was argued in these papers that the gradient catastrophe for dNLS is a source of the modulation instability for NLS.\par
The above two facts clearly indicate the importance of the integrable systems modeling certain effects of vortex filament dynamics and of their quasiclassical (or dispersionless) limit. \par
Moreover, the results of the papers \cite{LM,MHR}, showing that the cutoff $L/a$ is not a constant but strongly depends on curvature and torsion of filament, manifestly show the necessity to go beyond the approximation described by equations (\ref{Eqn-axial-approx}) and (\ref{Hirota}). Indeed, in general case, the parameters $\alpha_i$ in (\ref{eqn-axial}) are some functions of $K$, the vectors $\vec{P}$  and $\vec{Q}$ are certain superpositions of $\vec{t}$, $\vec{b}$, and $\vec{n}$ and then the induced velocity is of the form 
\begin{equation}
\label{X-generic-motion}
 \vec{X}_t =A(K,\tau) \vec{t}+B(K,\tau) \vec{n}+C(K,\tau) \vec{b}
\end{equation}
 where A,B,C are some functions. Equation (\ref{X-generic-motion}) implies certain intrinsic equations for $K$ and $\tau$ which, for particular choices of  A,B,C, are integrable systems \cite{NSW} different from (\ref{Hirota}).
 Relevance of these equations for the vortex filament dynamics is definitely of interest. \par
Study of the quasiclassical approximation for the intrinsic equations associated with the motion (\ref{X-generic-motion}) and their 
implications for the vortex filament dynamics is the main goal of the present paper. Quasiclassical intrinsic systems describe motions of 
vortex filament with curvature and torsion slowly varying along the filament and in the motion, i.e. they are systems of equations for 
$K=K(x,y)$, $\tau=\tau(x,y)$ where $x$ and $y$ are slow variables defined as $x=\epsilon s$, $y=\epsilon t$, $\epsilon \ll 1$ (see e.g. \cite{Zak}). 
For the Da Rios system (\ref{bin-intrinsic}) one has
\begin{equation}
\label{bin-intrinsic-disp}
\begin{split}
K_y&=-2K_x \tau -K \tau_x,\\
\tau_y &= KK_x -2\tau\tau_x 
\end{split}
\end{equation}
or, in terms of Riemann invariants $\beta=-\tau+iK$
\begin{equation}
 \label{bin-Riemann-inv}
\beta_y=\frac{1}{2}(3\beta+\overline{\beta})\beta_x.
\end{equation}
For the intrinsic system associated with equation (\ref{Hirota})  (see \cite{FM}) and others the corresponding quasiclassical equations are
 of the form 
\begin{equation}
\label{riem-inv-gen}
 \beta_y=\lambda(\beta,\overline{\beta}) \beta_x
\end{equation}
where the characteristic velocity is an appropriate function of $\beta$ and $\overline{\beta}$.\par
In this paper we will concentrate on the study of the equation 
\begin{equation}
 \label{NLS-mKdV-toda-riem-inv}
\beta_y=\left[\frac{1}{2}(3\beta+\overline{\beta})+\gamma_1\frac{3}{8}\left(5 \beta^2 +2\beta \overline{\beta} + \overline{\beta}^2 \right)
+\gamma_2 \frac{2}{\beta-\overline{\beta}} \right]\beta_x
\end{equation}
which is the the quasiclassical limit of the intrinsic equations corresponding to the evolution of the filament of the type (\ref{X-generic-motion}) with
\begin{equation}
  \label{NLS-mKdV-toda-X-motion}
 \begin{split}
A= \gamma_1 \frac{1}{2} K^2, \quad 
B= \gamma_1 K_x
+\gamma_2 \frac{1}{K}\left( \frac{\tau}{K^2+\tau^2}+K D \right)_x, \quad
C= K+ \gamma_1 K \tau +\gamma_2 \left( \frac{K}{K^2+\tau^2}+\tau D\right)
 \end{split}
\end{equation}
where $D=K\frac{\ln({\tau}/{K}+\sqrt{1+\tau^2/K^2})}{\left(K^2+\tau^2\right)^{3/2}}$ and $\gamma_1,\gamma_2$ being real constants. First two terms in r.h.s. of (\ref{NLS-mKdV-toda-riem-inv}) represents themselves the quasiclassical limit of equation (\ref{Hirota}) (see equations (3.76),(3.77)) from (\cite{FM} with $\gamma_1=\widetilde{W}$), while the last term is of different type. Thought the corresponding contribution to the velocity $v$ (\ref{X-generic-motion}) is quite complicated (\ref{NLS-mKdV-toda-X-motion}) the equation describing this flow separately is a very simple one. It is equivalent to the equation 
\begin{equation}
 \phi_{xx}+(e^\phi)_{yy}=0, \qquad \phi=K^2
\end{equation}
which is the dispersionless Toda (dToda) or the Boyer-Finley equation \cite{BF}-\cite{TT2}. dToda equation is integrable and plays an important role in various physical phenomena (see e.g. \cite{TT2}-\cite{KMM1}). It is known also that dToda and dNLS equations belong to the same dToda/dNLS hierarchy (see e.g. \cite{DGK,KMM1}). While the first two terms in r.h.s. of (\ref{NLS-mKdV-toda-riem-inv}) represent ``positive''  members the last (dToda) term is the first ``log-type'' term.  The presence of such dToda contribution into the induced velocity (\ref{X-generic-motion}), even if not completely justified in a model independent way, could be of phenomenological importance since it might describe nonlinear effects different from those captured by equation (\ref{Hirota}).\par
Ellipticity of equation (\ref{NLS-mKdV-toda-riem-inv}) or, in general, of (\ref{riem-inv-gen}) suggests to write it, following to \cite{Lav,Ahl,Boj}, as the nonlinear Beltrami equation
\begin{equation}
\label{nonl-Beltrami}
 \beta_{\overline{z}}=\frac{1+i\lambda(\beta,\overline{\beta})}{1-i\lambda(\beta,\overline{\beta})} \beta_z, \qquad z=x+iy.
\end{equation}
So, a solution of equation (\ref{NLS-mKdV-toda-riem-inv}) defines (see e.g. \cite{Lav,Ahl,Boj}) quasiconformal mapping if 
\begin{equation}
\left\vert
 \frac{1+i\lambda(\beta,\overline{\beta})}{1-i\lambda(\beta,\overline{\beta})}
\right\vert <1.
\end{equation}
In particular, any solution of the dDR system (\ref{bin-intrinsic-disp}) defines a quasiconformal mapping on the plane $z$. Quasiconformal mappings definitely play an important role in the vortex filament dynamics.\par
According to the result of papers \cite{EJLM}-\cite{JMcL}, equation (\ref{NLS-mKdV-toda-riem-inv}) well approximates the corresponding full equation of motion (equation at $\gamma_2=0$) until the derivatives $\beta_x$ and $\beta_y$ are not large. Analysis of behavior of solutions of equation (\ref{NLS-mKdV-toda-riem-inv}) near the point of gradient catastrophe, where $\beta_x$ and $\beta_y$ become unbounded, is the main subject of our study. Some results concerning dDR equation (equation \ref{NLS-mKdV-toda-riem-inv} with $\gamma_1=\gamma_2=0$) have been reported briefly in \cite{KOflutter}. Here, following the same approach, we will analyze the full equation (\ref{NLS-mKdV-toda-riem-inv}). We will see that the inclusion of these two terms essentially modify the behavior of vortex filament. \par
Our approach starts with the observation that the hodograph solutions of equation (\ref{NLS-mKdV-toda-riem-inv}) describe critical points $W_\beta=0$ of the function $W$ which obeys the Euler-Poisson-Darboux equation $E\left(\frac{1}{2},\frac{1}{2}\right)$, i.e.
\begin{equation}
 \label{EPD-NLS}
\left( \beta-\overline{\beta} \right) W_{\beta \overline{\beta}} = \frac{1}{2} \left( W_\beta - W_{\overline{\beta}} \right).
\end{equation}
 This fact essentially simplifies the analysis of the singular sector for equation (\ref{NLS-mKdV-toda-riem-inv}) associated with the gradient catastrophe. \par
Since $\beta_x,\beta_y \sim \frac{1}{W_{\beta\beta}}$ the points of gradient catastrophe belong to a singular sector $\mathcal{M}^{sing}$ in the space of solutions of the system (\ref{NLS-mKdV-toda-riem-inv}) for which $W_{\beta\beta}=0$.  Generic points of $\mathcal{M}^{sing}$ are characterized by the condition $W_{\beta\beta\beta} \neq 0$ while for deeper singular subsectors $\mathcal{M}^{sing}_n$, $n \geq 2$ all the derivatives of $W$ vanish up to $\partial_{\beta}^{n+1}W$. For these subsectors one has $2n$ constraints on all variables parameterizing $W$.
 In particular, generic gradient catastrophe happens in  a single point $x_0,y_0$. \par
Vortex filament exhibits a very particular behavior at the point of gradient catastrophe. While at a regular point (with bounded $K$, $\tau$,$K_x$, $\tau_x$) a curve modeling the filament always lies of one side of the rectifying plane (spanned  by the tangent vector $\vec{t}$ and the binormal vector $\vec{b}$), at the point of gradient catastrophe it oscillates fast around the rectifying plane. In addition, center of the osculating sphere (and its radius) blows up to infinity. Moreover, the induced velocity $\vec{V}=\vec{X}_t$ becomes unbounded at the point $x_0$, $y_0$. So, the gradient catastrophe for the system (\ref{NLS-mKdV-toda-riem-inv}) give rises to the fast oscillations which resemble the classical airfoil flutter (see e.g. \cite{MilT}). Numerical results \cite{CMM} show that such oscillations begin to expand along the filament \cite{KOflutter}.
Such a behavior of filament clearly indicates on its instability.\par
Analytically at the point of gradient catastrophe one has a catastrophe of certain type. Using the double scaling limit technique we show that at the generic point of gradient catastrophe (i.e. in the sector $\mathcal{M}^{sing}_1$) one has the elliptic umbilic catastrophe in terminology of R. Thom \cite{Thom} with the principal part given by $U^3-3 UV^2$. Deeper gradient catastrophes are associated with higher singularities. For example, in the sector $\mathcal{M}^{sing}_2$ the corresponding singularity is given by $X_9$ singularity from the V. Arnold list \cite{AGV} with the principal part of the form $U^4-6U^2V^2+V^4$.\par
At the point of gradient catastrophe the quasiclassical approximation becomes invalid and the system (\ref{NLS-mKdV-toda-riem-inv}) should be substituted by its full dispersive version  (by the intrinsic version of equation (\ref{Eqn-axial-approx}) at $\gamma_2=0$). First order corrections can be obtained by the double scaling technique together with an appropriate modification of the function $W^*(\beta^*,\obeta^*)$ to a functional $\int W_q(\beta^*,\obeta^*) dz^* d\overline{z}^*$ such that the equation ${W}_\beta=0$ for critical points are substituted by the Euler-Lagrange equation $\frac{\delta W_q}{\delta \beta^*}=0$ for $W_q$. It is shown that for generic point of gradient catastrophe (i.e. the sector 
$\mathcal{M}^{sing}_1$) the resulting equation for small corrections is equivalent to the Painlev\'e I equation
\begin{equation}
 \label{P-I}
\Omega_{\xi\xi}=6 \Omega^2-\xi.
\end{equation}
 The result of the paper \cite{DGK} allows us to conjecture that any generic solution of the Da Rios system (\ref{bin-intrinsic}) at the point of gradient catastrophe  behaves as $K=K_0+\varepsilon K^*$ and $\tau=\tau_0+\varepsilon \tau^*$ where $\beta^*=-\tau^*+iK^*$ is described by the tritronque\'e solution of the Painlev\'e I equation.\par
It is shown also that a class of double scaling regularizations of the higher gradient catastrophes associated with sectors $\mathcal{M}^{sing}_n$ is 
described by the equations 
\begin{equation}
 \label{P-I-n}
\Omega_{\xi\xi}= (n+2)!\Omega^{n+1}-\xi, \qquad n=2,3,\dots\ .
\end{equation}
The paper is organized as follows. In section 2 the relation of hodograph solutions with Euler-Poisson-Darboux equation is discussed. In section 3 it is shown that the hodograph solutions of dToda/dNLS equation describe critical points of a function. Hydrodynamic analogy and quasiconformal mappings for the governing equations are considered in section 4. Next section is devoted to an analysis of gradient catastrophe. Geometrical implications of gradient catastrophe and flutter of vortex filament are described in section 6. In section 7 it is shown that in the point of gradient catastrophe one has elliptic umbilic, $X_9$ and other singularities. Double scaling regularization of these catastrophes is analyzed in section 8.
\section{Hodograph solutions and Euler-Poisson-Darboux equation}
The dDR system (\ref{bin-Riemann-inv}), equation (\ref{NLS-mKdV-toda-riem-inv}) and general two-component elliptic quasilinear system written in terms of Riemann invariants, i.e. (\ref{riem-inv-gen}) are linearizable by the classical hodograph transformation $(x,y)\to (\mathrm{Re} \beta,\mathrm{Im} \beta)$ (see e.g. \cite{Wei,RY}). One has 
\begin{equation}
\label{hodo-gen}
 x_{\beta}+\overline{\lambda} y_\beta=0.
\end{equation}
For equation (\ref{riem-inv-gen}) in the form of nonlinear Beltrami equation (\ref{nonl-Beltrami}) the corresponding linearized equation is
\begin{equation}
 (1-i\overline{\lambda})z_{\beta}+(1+i\overline{\lambda})\overline{z}_{\beta}=0. 
\end{equation}
The compatibility condition for the system (\ref{hodo-gen}) is given by the second order equation
\begin{equation}
\label{hodo-EPD}
 (\lambda-\overline{\lambda})y_{\beta \overline{\beta}}
=\overline{\lambda}_{\overline{\beta}} y_\beta 
-{\lambda}_{\beta} y_{\overline{\beta}}
\end{equation}
solutions of which provide us with solutions of hodograph equations (\ref{hodo-gen}). In the hyperbolic case, i.e. for hydrodynamic type equations, equation of the type (\ref{hodo-EPD}) has been widely used for the construction of the solutions and an analysis of their properties (see e.g. \cite{Wei,RY,LL}).\par
In our elliptic case, equation of type (\ref{hodo-EPD}) written in term of variables $v=-\beta-\overline{\beta}$, $u=-(\beta-\overline{\beta})^2/4$ and $\beta$  defined as $y=\beta_v$, i.e. the equation
\begin{equation}
\label{lin-suleiman}
 u \beta_{uu}+2\beta_u+\alpha(u)\beta_{uv}=0, \qquad \alpha(u)>0
\end{equation}
 has been applied to the study of the system 
\begin{equation}
\label{suleiman-eqn}
 u_y+(uv)_x=0, \qquad v_y+vv_x-\alpha(u)u_x=0
\end{equation}
which appears in the theory of wave processes in unstable media within nonlinear geometric optics approximation \cite{GuSh,Shv}. At $\alpha(u)=1$ it is the quasiclassical limit of the focusing NLS equation (\ref{NLS}) or the dDR system (\ref{bin-intrinsic-disp}) for $u=K^2$, $v=2\tau$. In particular, in the paper \cite{KS}, the behavior of the solutions for the system (\ref{suleiman-eqn}) near the singular point $u=0$ of equation (\ref{lin-suleiman}) has been studied. For the hyperbolic version see also \cite{GaSu} \par
For the dDR system (\ref{bin-Riemann-inv}) equation (\ref{hodo-EPD}) is particularly simple one
\begin{equation}
\label{hodo-EPD-NLS}
 (\beta-\overline{\beta})y_{\beta\overline{\beta}}=\frac{3}{2}(y_\beta-y_{\overline{\beta}}).
\end{equation}
It is the Euler-Poisson-Darboux (EPD) equation of type $E\left(\frac{3}{2},\frac{3}{2}\right)$ (see e.g. \cite{Dar}). For the dToda equation
 $\beta_y=\frac{1}{\beta-\overline{\beta}} \beta_x$, equation (\ref{hodo-EPD}) is of the form  
\begin{equation}
\label{EPD-Toda}
 (\beta-\overline{\beta})y_{\beta\overline{\beta}}=\frac{1}{2}(y_\beta+y_{\overline{\beta}}).
\end{equation}
that is again the simple EPD equation of the type $E\left(\frac{1}{2},-\frac{1}{2}\right)$.\par
In contrast, for equation (\ref{hodo-gen}) with $\lambda=\frac{3}{8} (5\beta^2 +2\beta\overline{\beta}+\overline{\beta}^2)$ which corresponds to the second term in (\ref{NLS-mKdV-toda-riem-inv})  and to the quasiclassical limit of the first higher NLS equation (see formula (\ref{Hirota})), equation (\ref{hodo-EPD}) is
\begin{equation}
\label{nonEPD-mKdV}
 (\beta^2-\obeta^2)y_{\beta \obeta}= 2(5 \obeta +2\beta) y_\beta - 2 (5 \beta +2 \obeta) y_{\obeta}.
\end{equation}
For higher dNLS equations this equation is more and more complicated and hence an analysis similar to that performed in \cite{Shv,GaSu} becomes much more involved.\par
A way to simplify drastically equations of type (\ref{hodo-EPD}) associated with equation (\ref{hodo-EPD}) can be extracted from the Tsarev's
\cite{Tsa} discussion of the generalized hodograph equations. He noted that it is convenient to rewrite equation (\ref{hodo-gen}) as
\begin{equation}
\label{hodo-gen-Tsarev}
 (x+\lambda y)_{\obeta} -y \lambda_{\obeta}=0.
\end{equation}
Denoting $x+\lambda y=\omega(\beta,\obeta)$, one observes that $y=\frac{\omega-\overline{\omega}}{\lambda-\overline{\lambda}}$ and equation (\ref{hodo-gen-Tsarev}) assumes the form
\begin{equation}
\label{hodo-constr-Tsarev}
 \frac{\omega_{\obeta}}{\omega-\overline{\omega}}=\frac{\lambda_{\obeta}}{\lambda-\overline{\lambda}}.
\end{equation}
So, equation (\ref{hodo-gen-Tsarev}) and, hence, the hodograph equation (\ref{hodo-gen}) takes the form
\begin{equation}
\label{hodo-new-Tsarev}
 x+\lambda(\beta,\obeta)y-\omega(\beta,\obeta)=0
\end{equation}
where $\omega(\beta,\obeta)$ is any function obeying the constraint (\ref{hodo-constr-Tsarev}).  In the form (\ref{hodo-new-Tsarev}) the hodograph equations can be formulated also for multi-component hydrodynamic type systems \cite{Tsa}. \par
An observation here is that for the dDR system ({\ref{bin-intrinsic-disp}}) and equation (\ref{bin-Riemann-inv}) one has
\begin{equation}
\label{lambda-eqn}
 \frac{\lambda_{\obeta}}{\lambda-\overline{\lambda}}=\frac{1}{2(\beta-\obeta)}
\end{equation}
and, hence,
\begin{equation}
\label{omega-eqn}
 \frac{\omega_{\obeta}}{\omega-\overline{\omega}}=\frac{1}{2(\beta-\obeta)}.
\end{equation}
Furthermore, these equations imply that 
\begin{equation}
 \lambda=\widetilde{W}_\beta, \qquad \omega=\widetilde{\widetilde{W}}_\beta 
\end{equation}
where $\widetilde{W}(\beta,\obeta)$ and $\widetilde{\widetilde{W}}(\beta,\obeta)$ are certain real valued functions.
For instance, for equation (\ref{NLS-mKdV-toda-riem-inv}) 
\begin{equation}
 \label{gen-W}
\widetilde{W}=y\left[
\frac{1}{8}(3\beta^2+2\overline{\beta}\beta+3\overline{\beta}^2)+
\gamma_1\frac{1}{16}(5\beta^3+3\beta^2\obeta+3\obeta^2\beta+5\obeta^3)+
\gamma_2 \ln\left(\frac{\beta-\obeta}{2i}\right)
\right].
\end{equation}
 In virtue of (\ref{lambda-eqn}) and (\ref{omega-eqn}) the functions $\widetilde{W}$ and $\widetilde{\widetilde{W}}$ obey the EPD equation $E\left(\frac{1}{2},\frac{1}{2}\right)$
\begin{equation}
 \begin{split}
  &(\beta-\obeta) \widetilde{W}_{\beta \obeta}=\frac{1}{2} \left(\widetilde{W}_\beta-\widetilde{W}_{\obeta}\right), \\
&(\beta-\obeta) \widetilde{\widetilde{W}}_{\beta \obeta}=\frac{1}{2} \left(\widetilde{\widetilde{W}}_\beta-\widetilde{\widetilde{W}}_{\obeta}\right).  
\end{split}
\end{equation}
So, the hodograph equation (\ref{hodo-gen-Tsarev}) is of the form
\begin{equation}
\label{hodo-fin}
 x+y \widetilde{W}_\beta -\widetilde{\widetilde{W}}=0
\end{equation}
where ${\widetilde{W}}$ and $\widetilde{\widetilde{W}}$ are solutions of the same EPD equation $E\left(\frac{1}{2},\frac{1}{2}\right)$.
It remains to note that 
\begin{equation}
 y=\frac{\widetilde{\widetilde{W}}_\beta-\widetilde{\widetilde{W}}_{\obeta}}{\widetilde{W}_\beta-\widetilde{W}_{\obeta}}
=\frac{\widetilde{\widetilde{W}}_{\beta \obeta}}{\widetilde{W}_{\beta \obeta}}
\end{equation}
is a solution of the equation (\ref{hodo-EPD}) with $\lambda = \widetilde{W}_\beta$.\par
Thus, in contrast to the usual approach in which hodograph solutions of the commuting quasilinear systems contributing into the r.h.s. of equation (\ref{NLS-mKdV-toda-riem-inv}) can be found via different linear second order equations (see equations (\ref{EPD-NLS}),(\ref{EPD-Toda}), (\ref{nonEPD-mKdV})), in the approach described above all of them can be build via solutions of the same and simple EPD equation
$E\left(\frac{1}{2},\frac{1}{2}\right)$. \par
EPD equations of different types are well studied and have number of remarkable properties (see e.g. \cite{Dar}). 
Their relevance for the theory of hydrodynamic type equations and Whitham equations has been understood in the papers \cite{KuSh}-\cite{KMM2}.
\section{Quasiclassical intrinsic equations for dToda/dNLS hierarchy and critical points of functions}
In addition to the observations made in the previous section, one readily concludes that the hodograph equation (\ref{hodo-fin}) can be represented as
\begin{equation}
\label{crit-eqn}
 W_\beta=0
\end{equation}
where
\begin{equation}
\label{W}
 W=\frac{1}{2}x(\beta+\obeta)+y\widetilde{W}-\widetilde{\widetilde{W}}
\end{equation}
and the function $W$ obeys the EPD $E\left(\frac{1}{2},\frac{1}{2}\right)$ equation 
\begin{equation}
\label{EPD-W}
 (\beta-\obeta)W_{\beta\obeta}=\frac{1}{2}\left(W_{\beta}-W_{\obeta}\right).
\end{equation}
Thus, one has
\begin{prop}
 Hodograph solutions of equations (\ref{NLS-mKdV-toda-riem-inv}) describes critical points (\ref{crit-eqn}) of the function $W$ defined in (\ref{W}) which obeys the EPD $E\left(\frac{1}{2},\frac{1}{2}\right)$ equation.
\end{prop}
In the hyperbolic case, i.e. for the quasiclassical shallow water equation (or one-layer Benney system) this fact has been observed in \cite{KMM} (see also \cite{KMM2}).\par
So, in order to construct a solution of equation (\ref{NLS-mKdV-toda-riem-inv}) one takes a function $W$ of the form (\ref{W}) with function
${\widetilde{W}}$ given by (\ref{gen-W}), arbitrary functions $\widetilde{\widetilde{W}}$ solving EPD $E\left(\frac{1}{2},\frac{1}{2}\right)$ equation  and finds its critical points solving equation (\ref{crit-eqn}).  The class of solutions obtained in this way is controlled by the class of solutions of the EPD $E\left(\frac{1}{2},\frac{1}{2}\right)$ equation. Various classes of solutions of this equation has been described a century ago (see e.g. \cite{Dar}) as well as in recent papers concerning the application of the EPD equation in Whitham theory. For example, a class of solutions of the EPD $E(\epsilon,\epsilon)$ equation is given by
\begin{equation}
\label{e-W}
 W=\oint_S \frac{d\lambda}{2\pi i} u(\lambda,t)\left(1-\frac{\beta}{\lambda}\right)^{-\epsilon}\left(1-\frac{\obeta}{\lambda}\right)^{-\epsilon}
\end{equation}
where $S$ is the positively oriented circle of large radius and $u(\lambda,t):= \sum_{k=0}^N \lambda^kt_k$ is an arbitrary polynomial of the complex variable $\lambda$ parameterized by the constants $t_k$ (see e.g. \cite{KMM,KMM2}). In particular, the function $\widetilde{W}$ (\ref{W}) with $\gamma_2=0$ is given by (\ref{e-W}) with 
$\epsilon=\frac{1}{2}$ and $u=x+y(\lambda+\gamma_1\lambda^2)$. Parameters $t_k$  in $u$ and $W$ can be viewed as the variables parameterizing a class of initial data for equation (\ref{NLS-mKdV-toda-riem-inv}).\par
In terms of curvature and torsion the EPD equation (\ref{EPD-W}) is of the form
\begin{equation}
 K(W_{KK}+W_{\tau\tau})+W_K=0
\end{equation}
that is the axisymmetric three dimensional  Laplace equation studied by Beltrami \cite{Bel} (see also \cite{Wei}). Here $K$ and $\tau$ play the role of the radial and axial coordinates, respectively, in the cylindrical system of coordinates.\par
Function $W$ (\ref{W}) is 
\begin{equation}
\label{W-tot-KT} 
W=-x\tau +y \left[ \left(\tau^2-\frac{1}{2}K^2\right)+\gamma_1\left(-\tau^3+\frac{3}{2}K\tau^2\right)+\gamma_2 \ln K \right]-\widetilde{\widetilde{W}}
\end{equation}
 and hodograph equations are given by 
\begin{equation}
\label{crit-eqn-KT}
\begin{split} 
& W_K=y \left[-K+\gamma_1\frac{3}{2}\tau^2+\frac{\gamma_2}{K} \right]-\widetilde{\widetilde{W}}_K=0, \\
&W_\tau=-x+y \left[ 2\tau + \gamma_1  \left( -3\tau^2 + 3 K \tau \right) \right]-\widetilde{\widetilde{W}}_\tau=0.
\end{split}
\end{equation}
The corresponding  system of quasilinear PDEs in terms of $K$ and $\tau$ is of the form
\begin{equation}
\label{KT-sys} 
\begin{split}
K_y +& \left[ 2\tau-\gamma_1\left( 3\tau^2-\frac{3}{2}K^2\right)  \right] K_x+\left[ K-3\gamma_1 \tau K -\frac{\gamma_2}{K}\right] \tau_x =0,
 \\
  \tau_y -& \left[ K-3\gamma_1 \tau K -\frac{\gamma_2}{K}\right] K_x +\left[2\tau-\gamma_1\left( 3\tau^2-\frac{3}{2}K^2\right)  \right] \tau_x =0.            
   \end{split}
\end{equation}
At $\gamma_2=0$ it coincides with the dispersionless limit of the equations given in \cite{FM}.
We note that the contribution from dToda terms (with $\gamma_2 \neq 0$) are quite different from those of dispersionless Hirota equation ((\ref{KT-sys}) with $\gamma_2=0$) considered in \cite{FM}. Formally at small $K$, the last terms in (\ref{KT-sys}) becomes dominant. Essential difference between the dDR and dHirota type evolution and dToda contribution is manifested by the expression (\ref{NLS-mKdV-toda-X-motion}) of the velocity of filament. The former are small for filaments close to a straight line (small $K$) and becomes larger for more curved filaments. In contrast, the dToda type contribution to the induced velocity (proportional to $\gamma_2$) is smaller for larger $K$ and increase with decreasing $K$. Since in the approximation under consideration always $K \ll 1 / a$ the dToda type contribution might be relevant for the description of some nonlinear and nonlocal effects appearing for slightly curved vortex filaments.\par
In virtue of (\ref{EPD-W}) at the critical points of the function $W$ given by   (\ref{crit-eqn}) or (\ref{crit-eqn-KT}) one has
\begin{equation}
\label{EPD-crit}
 (\beta-\obeta)W_{\beta\obeta}=0
\end{equation}
or
\begin{equation}
\label{EPD-crit-KT}
 K(W_{KK}+W_{\tau\tau})=0.
\end{equation}
The point $K=0$ is a singular point for these equations. Analysis of the behaviors of solutions of these equations near this point can be performed in a way similar to that discussed in \cite{KS}.\par
Geometrically the case $K=0$ corresponds to a straight line. In this paper we will be interested in properties of curved vortex filament. In the rest of the paper we will assume that $K \neq 0$. So, equations (\ref{EPD-crit}) and (\ref{EPD-crit-KT}) take the form
\begin{equation}
\label{WBBb}
 W_{\beta\obeta}=0
\end{equation}
and
\begin{equation}
 W_{KK}+W_{\tau\tau}=0.
\end{equation}
We will see that these conditions simplify essentially an analysis of singular sector for equation (\ref{NLS-mKdV-toda-riem-inv}).
\section{Ellipticity of governing equations: hydrodynamic analogy and quasiconformal mappings}
Ellipticity of the dDR system (\ref{bin-intrinsic-disp}) and more general system (\ref{KT-sys}) means that they have properties drastically different from those of the hyperbolic version of the dDR system  (or the system (\ref{suleiman-eqn}) with $\alpha=-1$), that is the classical shallow water equation (see e.g. \cite{RY}). The latter has standard Riemann wave solutions  while in the elliptic case the Cauchy problem is ill posed. In spite of such a difference a representation of the elliptic systems in the form of hydrodynamic type systems, but with negative pressure, has been widely used in the analysis of their local properties (see \cite{Bet,GuSh,Shv,FM,EGKK}).\par
Following this line one can rewrite the system (\ref{KT-sys}) in terms of $\rho=K^2$ and 
$v=2\left(1-\frac{\gamma_2}{K^2}\right)\tau-3\gamma_1\left(\tau^2-\frac{1}{4}K^2\right)$ as
\begin{equation}
\label{Euler}
 \begin{split}
  & \rho_y+(v \rho)_x=0, \\
&\rho(v_y+vv_x)+P_x=0
 \end{split}
\end{equation}
where 
\begin{equation}
\label{Pressure}
P=-\frac{1}{2} \left(1-3\gamma_1\tau \right)^2\rho^{2}
+\frac{3}{16}\gamma_1^2  \rho^3 
-\gamma_1\gamma_2 \tau \left( 3\rho +8\tau^3 \right)
+2 \gamma_2 \left(  \rho + 2 \tau^2\right)
-\gamma_2^2 \left(4\frac{\tau^2}{\rho}+\ln \rho \right).
\end{equation}
Here $\rho$, $v$, and $P$ play the roles of density, velocity and pressure. Note that the system considered in \cite{FM} is not a system of quasi-linear equations. It contains the terms nonlinear in first order derivatives in addition to those quasi-linear ones in the system (\ref{KT-sys}) with $\gamma_2=0$. Moreover our definition of the velocity and pressure is slightly different from those from \cite{FM}.
Consequently, the system (\ref{Euler}) is composed exactly by the continuity and Euler equations in contrast to the system (3.79)-(3.80) from \cite{FM}.\par 
Properties of the system (\ref{Euler}) are fully determined by the properties of the pressure (\ref{Pressure}). Within LIA the pressure $P=-\rho^2/2$ is always negative and that causes the well-known instability of vortex filament. Inclusion of the Hirota and dToda contributions changes the situation drastically. For large curvature axial flow contributions dominate and the sign of the pressure is different for different values of $K$ and $\tau$. For instance, if $K>2\sqrt{6} |\tau|$ then the pressure P for large $K$ and $\tau$ is positive, while in the opposite case it is negative. Then, for the points with small $\tau$ (near flattening points) one has stabilizing effect ($P>0$) while, with large torsion $\tau \sim a^{-1}$ dHirota contribution (i.e. axial flow contribution) always destabilize the filament (recall that $K \ll a^{-1}$).\par
For small $K$ (small density) pure dToda contribution equal to  
$-\gamma_2^2 \left(\ln \rho+ {\tau^2}/{\rho} \right)$  dominates the pressure. 
So, for small $K$ the sign of the pressure coincides with the sign of 
$4\tau^2+2K^2\ln K$. Hence, once again one has a threshold dividing the regimes with positive and negative pressure and consequently stabilizing and destabilizing effects of contributions different from the original LIA model of Da Rios. If one considers the contribution only from axial flow ($\gamma_1 \neq 0,\ \gamma_2=0$) the pressure is
\begin{equation}
 P=-\frac{1}{2}(1-3\gamma_1\tau)^2 \rho^2 +\frac{3}{16}\gamma_1^2 \rho^3.
\end{equation}
$P$ vanishes for $K$ and $\tau$ connected by the relation $3 \gamma_1^2 K^6=8 (1-3\gamma_1\tau)^2K^4$, while it is positive for vortex filaments with curvature and torsion obeying the condition
\begin{equation}
 K>\sqrt{\frac{8}{3}} \left\vert \frac{1}{\gamma_1}- 3 \tau \right\vert
\end{equation}
 and it is negative in the opposite case.\par
This discussion shows that stability or instability of vortex filaments strongly depends on their geometry (relations between curvature and torsion). \par
There is another way to formulate and emphasize an ellipticity of governing equations.  
The dDR system (\ref{bin-Riemann-inv}) and more generally system (\ref{riem-inv-gen}) are elliptic systems of quasilinear equations and, hence, under certain conditions their solutions define quasiconformal mappings on the plane (see e.g. \cite{Lav,Ahl,Boj}). Introducing the independent variable $z=x+iy$, one can rewrite general equation (\ref{hodo-gen}) in the form
\begin{equation}
 \beta_{\overline{z}}=\frac{1+i\lambda(\beta,\overline{\beta})}{1-i\lambda(\beta,\overline{\beta})} \beta_z.
\end{equation}
It is the nonlinear Beltrami equation \cite{Lav,Ahl,Boj} solution of which defines a quasiconformal mapping $(z,\overline{\zeta})\to(\beta,\obeta)$ iff the dilation coefficient $\mu=\frac{1+i\lambda(\beta,\overline{\beta})}{1-i\lambda(\beta,\overline{\beta})}$ obeys the constraint $|\mu|<1$. This condition, i.e.
\begin{equation}
 \frac{|1+i\lambda(\beta,\overline{\beta})|}{|1-i\lambda(\beta,\overline{\beta})|}<1,
\end{equation}
imposes constraints on $\beta$ and, hence, on $K$ and $\tau$. For the dDR system (\ref{bin-intrinsic-disp}), $\lambda=\frac{1}{2}(3\beta+\obeta)=-2\tau+iK$, and one has
\begin{equation}
 |\mu|^2=\frac{4\tau^2+(1-K)^2}{4\tau^2+(1+K)^2}.
\end{equation}
 Since $K \geq 0$ one always has $|\mu|<1$ for any nonstraight filament. \par
So, the slow motions of nonstraight vortex filaments within original LIA are described by quasiconformal mappings on the plane defined by the equation
\begin{equation}
 \beta_{\overline{z}}=\frac{2+i(3\beta+\obeta)}{2-i(3\beta+\obeta)} \beta_z.
\end{equation}
  For the generalization of the LIA given by equation (\ref{Hirota}) or (\ref{NLS-mKdV-toda-riem-inv}) situation is more complicated. For equation (\ref{NLS-mKdV-toda-riem-inv}) with $\gamma_2=0$ one has
\begin{equation}
 |\mu|^2=\frac{8+i[4(3\beta+\obeta)+3\gamma_1(5\beta^2+2\beta\obeta+\obeta^2)]}{8-i[4(3\beta+\obeta)+3\gamma_1(5\beta^2+2\beta\obeta+\obeta^2)]}
\end{equation}
and the condition $|\mu<1|$ is verified if $K$ and $\tau$ and parameter $\gamma_1$ obey the constraint
\begin{equation}
 \begin{split}
  K \neq 0, \qquad \frac{1-3\gamma_1 \tau}{(2+2K)^2+(4\tau-6\gamma_1\tau)^2+3\gamma_1K(3\gamma_1K-8\tau)} > 0. 
 \end{split}
\end{equation}
So, only part of the vortex filament configurations with axial flow is associated with the quasiconformal mappings, a sufficient condition being $\gamma_1 \tau <0$. 
For the complete flow (\ref{NLS-mKdV-toda-riem-inv}) the constraint on $K,\tau,\gamma_1$ and $\gamma_2$ is more complicated. It is of interest to note that for pure dToda equation 
\begin{equation}
\label{riem-inv-dToda}
 \beta_y=\frac{2\gamma_2}{\beta-\obeta}\beta_x
\end{equation}
corresponding to the last term in (\ref{NLS-mKdV-toda-riem-inv}), one has
\begin{equation}
 |\mu|=\frac{K+\gamma_2}{K-\gamma_2}.
\end{equation}
So, solutions of the dToda equation define a quasiconformal mapping iff $\gamma_2<0$. At $\gamma_2=-1$ equation (\ref{riem-inv-dToda}) is equivalent to the equation 
\begin{equation}
\label{BF-eqn}
 \phi_{xx}+(e^\phi)_{yy}=0 ,\qquad \phi=2\ln K
\end{equation}
and $|\mu|=\frac{K-1}{K+1}$ and, hence, any solution of the dToda equation (\ref{BF-eqn}) defines a quasiconformal mapping on the plane.\par
The results of the previous sections indicate also a relation between certain class of quasiconformal mappings on the plane and EPD equation.
Indeed, solutions of equation (\ref{crit-eqn}) with $\lambda=\widetilde{W}_\beta$, i.e. solutions of the Beltrami equation
\begin{equation}
\label{Beltrami-W}
 \beta_{\overline{z}}=\frac{1+i\widetilde{W}_\beta}{1-i\widetilde{W}_\beta}\beta_{z},
\end{equation}
are critical points $W_\beta=0$ of the function $W=\frac{1}{2}x(\beta+\obeta)+y\widetilde{W}-\widetilde{\widetilde{W}}$ which obeys the EPD
 $E\left(\frac{1}{2},\frac{1}{2}\right)$ equation. This means that the class of quasiconformal mappings on the plane defined by equation (\ref{Beltrami-W}) with arbitrary real-valued function  $\widetilde{W}(\beta,\obeta)$ obeying EPD
 $E\left(\frac{1}{2},\frac{1}{2}\right)$ equation describe critical points of the function 
\begin{equation}
\label{Wz}
 W=z(\beta+\obeta-2i\widetilde{W})+\overline{z}(\beta+\obeta+2i\widetilde{W})+\widetilde{\widetilde{W}}
\end{equation}
where $\widetilde{\widetilde{W}}$ is an arbitrary solution of the EPD $E\left(\frac{1}{2},\frac{1}{2}\right)$ equation. Inverse statement is valid too: critical points of the function (\ref{Wz}) obeying the EPD $E\left(\frac{1}{2},\frac{1}{2}\right)$ equation and constraint 
$\left| \frac{1+i\widetilde{W}_\beta}{1-i\widetilde{W}_\beta}\right|<1$ are described by quasiconformal mappings defined by equation (\ref{Wz}). Similar results are valid also for hydrodynamic type system of $\epsilon$-type and other types. \par
The interrelation between quasiconformal mappings and vortex filament dynamics will be studied elsewhere.
\section{Gradient catastrophe}
Hodograph  equations $W_\beta=0$ and $W_{\overline{\beta}}=0$ provide us with a single-valued solution of the dDR system (\ref{bin-Riemann-inv}) if the standard condition
\begin{equation} 
 \Delta=\mathrm{det} \left( \begin{array}{cc}  W_{\beta\beta} & W_{\beta\obeta} \\ W_{\obeta\beta} & W_{\obeta\obeta} \end{array} \right) \neq 0
\end{equation}
is satisfied. In virtue of the EPD equation (\ref{EPD-W}) on the solution of the systems (\ref{bin-intrinsic-disp}) and (\ref{bin-Riemann-inv})
with $K \neq 0$  the function $W$ obeys the equation $W_{\beta \obeta}=0$. Hence, at $K \neq 0$ 
\begin{equation}
 \Delta = |W_{\beta\beta}|^2.
\end{equation}
  Consequently, the regular sector of equations (\ref{bin-intrinsic-disp})  and (\ref{bin-Riemann-inv}) for which hodograph equations are uniquely solvable is defined by the conditions
\begin{equation}
 W_\beta=0,\qquad W_{\beta\beta}\neq 0.
\end{equation}
In this sector both $\beta$ and $\beta_x,\beta_y$ are bounded. Indeed, differentiating the hodograph equation $W_\beta=0$ with respect to $x$ and $y$ and using the equation (\ref{WBBb}), one gets
\begin{equation}
\label{bxby}
 \beta_x=-\frac{1}{2W_{\beta\beta}}, \qquad \beta_y=-\frac{\widetilde{W}_\beta}{2W_{\beta\beta}}.
\end{equation}
\par Gradient catastrophe sector or singular sector for equations (\ref{NLS-mKdV-toda-riem-inv}) is composed by those of their solutions for which  $\beta$ (i.e. curvature $K$ and torsion $\tau$) remain bounded while derivatives $\beta_x$, $\beta_y$ explode (for standard definition see e.g. \cite{RY}). Thus, for the gradient catastrophe sector 
\begin{equation}
\label{grad-cat-sec}
W_\beta=0, \qquad W_{\beta\beta}=0.
\end{equation}
Condition of solvability of these equations is $W_{\beta\beta\beta}\neq 0$ and, hence, generic singular sector is characterized by the conditions
\begin{equation}
\label{grad-cat-sec-cond}
W_\beta=0, \qquad W_{\beta\beta}=0, \qquad W_{\beta\beta\beta}\neq 0.
\end{equation}
In terms of curvature and torsion these conditions at $K \neq 0$ have the form
\begin{equation}
 \label{grad-cat-sec-cond-KT}
W_K=0, \quad W_\tau=0, \quad W_{K\tau}=0,\quad W_{KKK}\neq 0, \quad W_{\tau\tau\tau}\neq0.  
\end{equation}
Equations $W_{KK}=W_{\tau\tau}=0$, $W_{KK\tau} \neq 0$, $W_{K\tau\tau} \neq 0$ are the consequences of (\ref{grad-cat-sec-cond-KT}).\par
Critical points of functions which obey the condition (\ref{grad-cat-sec}) are called degenerate critical points (see e.g. \cite{AGV}).
Degenerate critical points are typical for families of functions parameterized by several variables \cite{AGV}. In our case the family of functions $W$ (\ref{W}) is parameterized by $x,y,\gamma_1,\gamma_2$ and other variables defining initial values of $\beta$ encoded in $\widetilde{\widetilde{W}}$.\par
In the generic situation the last condition in (\ref{grad-cat-sec-cond}) allow us to solve equations $W_{\beta \beta}=0$, $W_{\obeta\obeta}=0$ with respect to $\beta$ and $\obeta$, i.e. K and $\tau$. Substituting these expressions for $K$ and $\tau$ into the first equation (\ref{grad-cat-sec-cond}), one gets two relations
\begin{equation}
\label{crit-point-f}
 f_1(x,y;\gamma_1,\gamma_2)=0,\qquad  f_2(x,y;\gamma_1,\gamma_2)=0
\end{equation}
for equation (\ref{NLS-mKdV-toda-riem-inv}). Thus, the generic gradient catastrophe for equation (\ref{NLS-mKdV-toda-riem-inv}) 
is characterized by two constraints on the independent variables $x,y$ and parameters $\gamma_1,\gamma_2$. For fixed $\gamma_1$ and $\gamma_2$ 
the gradient catastrophe for equation (\ref{NLS-mKdV-toda-riem-inv}) occurs in a single point $(x_0,y_0)$ for 
given $\widetilde{\widetilde{W}}(\beta,\obeta)$, i.e. for given initial data for $K$ and $\tau$. For the dNLS equation $\gamma_1=\gamma_2=0$ 
this fact has been observed within another approach for the first time in \cite{DGK}. In a different way it was derived in \cite{KMM2}.\par
For the quasiclassical counterpart of the Hirota equation (\ref{Hirota}), i.e. equation   (\ref{NLS-mKdV-toda-riem-inv}) with $\gamma_2=0$, 
one has a one-parametric family of points of gradient catastrophes for given initial data. Recall that the parameter $\gamma_1$  
in equation (\ref{Hirota}) and  (\ref{NLS-mKdV-toda-riem-inv}) encodes the matching data conditions for inner and outer expansions for the 
vortex filament with axial flow \cite{MS}-\cite{FM}. Thus, the meaning of a true physical ``time'' of the gradient catastrophe should be assigned to the
 minimal value $y_0(\gamma_1)$ of the solutions $y(\gamma_1)$ of the equation (\ref{crit-point-f}) at $\gamma_2=0$. It happens in the 
corresponding point $x_0(\gamma_1)$ of the filament.\par
For equation (\ref{NLS-mKdV-toda-riem-inv}) and for the class of solutions of the dDR equation corresponding to a family of initial 
``data'' $\widetilde{\widetilde{W}}$ parameterized by additional variables, one can go deeper into the singular sector. Indeed it may 
happens that in addition to the conditions (\ref{grad-cat-sec}) in the singular sector the condition $W_{\beta\beta\beta}=0 $ is verified. For solutions for which $W_{\beta\beta\beta\beta}\neq 0$ one has a subsector defined by
\begin{equation}
\label{grad-cat-sec2-cond}
 W_\beta=0, \quad W_{\beta \beta}=0, \quad W_{\beta \beta \beta}=0, \quad W_{\beta \beta \beta \beta} \neq 0.
\end{equation}
Continuing this procedure one can show that the singular sector $\mathcal{M}^{sing}$ of the dDR equation or equations 
of type (\ref{NLS-mKdV-toda-riem-inv}) has the following structure \cite{KMM,KMM2}
\begin{equation}
 \mathcal{M}^{sing} =\bigcup_{n \geq 1} \mathcal{M}^{sing}_n
\end{equation}
where the subsector $\mathcal{M}^{sing}_n$ is defined as
\begin{equation}
 \mathcal{M}^{sing}_n=\{ (z,y;\gamma_1;\beta,\obeta) \in \mathcal{M}^{sing} : \frac{\partial^k W}{\partial \beta^k}=0,\ k=1,2,\dots,n+1;
\frac{\partial^{n+2} W}{\partial \beta^{n+2}}\neq0 \}.
\end{equation}
Solutions belonging to $\mathcal{M}^{sing}_n$ are defined on a subspace of codimension $2n$ in the space of the variables 
$x,y;\gamma_1,\gamma_2$ given by equations
\begin{equation}
\label{crit-point-f-n}
 f_k(x,y;\gamma_1,\gamma_2)=0,\qquad  k=1,2,\dots,2n.
\end{equation}
In order to have nonempty singular subsector $\mathcal{M}^{sing}$ one should have class of solutions characterized at least by 
$2n$ parameters. So, for equation (\ref{NLS-mKdV-toda-riem-inv}) with fixed ``initial data'' $\widetilde{W}(\beta,\obeta)$ one may have only 
two nonempty sectors $\mathcal{M}^{sing}_1$ and $\mathcal{M}^{sing}_2$. For appropriate family of the initial data one can have gradient 
catastrophe of arbitrary deepness.
\section{Flutter of filament}
A gradient catastrophe for the system (\ref{KT-sys}) means a special behavior of a filament at the point $x_0$ at time $y_0$. Using the formulae (\ref{X-generic-motion}) and (\ref{NLS-mKdV-toda-X-motion}), one observes that the induced velocity $\vec{V}$ has a particular behavior at the point of gradient catastrophe. Within pure LIA $\gamma_1=\gamma_2=0$, $\vec{V}=K\vec{b}$ and, hence, velocity remain bounded at the point $x_0$. However, the acceleration given by
\begin{equation}
 \vec{A}=\vec{V}_t=\left[ (-2K_s\tau - K \tau_s) \vec{b} -KK_s \vec{t} -\tau^2 \vec{n}\right]
\end{equation}
   explodes at $x_0$.\par
Axial flow ($\gamma_1 \neq 0$) and dToda contribution ($\gamma_2 \neq 0$) qualitatively change this behavior. Indeed, the formula (\ref{NLS-mKdV-toda-X-motion}) shows that at the point of gradient catastrophe the coefficient $B$, i.e. the normal component of the velocity becomes very large both for dHirota and dToda contributions. So, in the point of gradient catastrophe the local rate of self-stretching, proportional to $\vec{X}_t \cdot \vec{n} = B$ (see e.g. \cite{KM,NSW}) becomes very large. Note that such an effect is absent in pure dDR dynamics.
Sign of normal component depends on the sign of $K_x$ and $\tau_x$ and, hence, at the point $x_0$ the filament begins to oscillate changing fast the sign of normal components of velocity. \par
\begin{figure}[!ht]
\centering
\includegraphics[width=8cm, height=5cm]{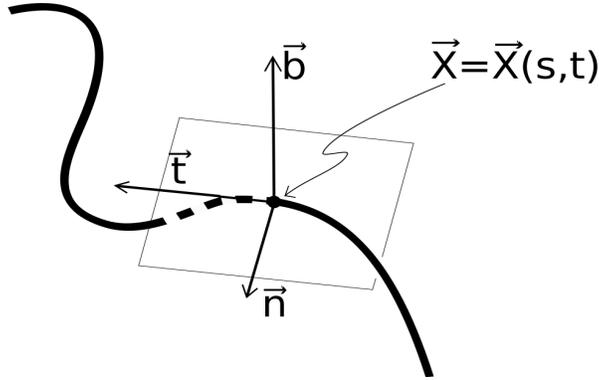}
\caption{Generic curve with $\tau>0$ at a regular point.}
\label{fig-vorline}
\end{figure}
\begin{figure}[!ht]
\centering
\includegraphics[width=15cm, height=5cm]{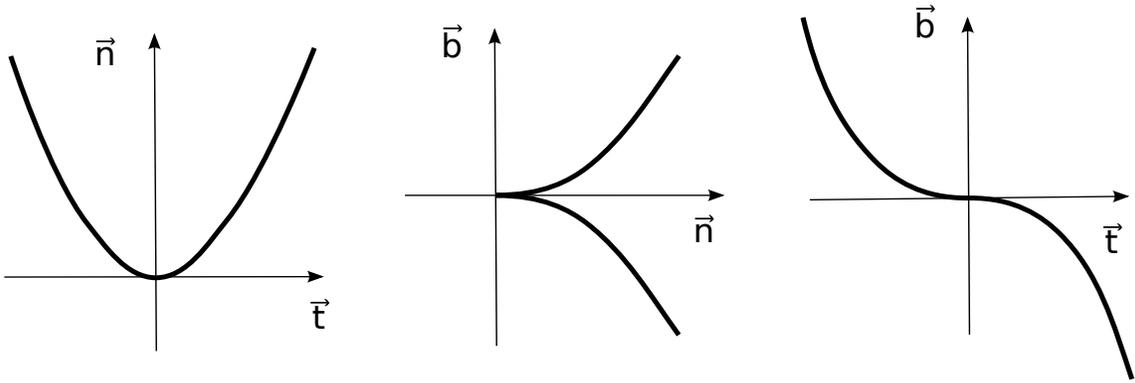}
\caption{The figure show a curve with $\tau>0$. At $\tau<0$ one should reflect the curve with respect to the axis $\vec{t}$ in the last figure.}
\label{fig-pcc}
\end{figure}

Local intrinsic consideration reveals a geometric origin and nature of such oscillations of the filament at the point of gradient catastrophe. 
It is well  known, that the coordinates of a curve in a neighborhood of a point in the reference system formed by the tangent vector $\vec{t}$, normal $\vec{n}$ and binormal $\vec{b}$ and with origin in the point (see e.g. formula (6.7) from \cite{Eis} and figure \ref{fig-vorline}) are
\begin{equation}
 \label{xxx}
\begin{split}
 & x_1=s-\frac{K^2}{6}s^3+\dots, \\
 & x_2=\frac{K}{2}s^2-\frac{K_s}{6}s^3+\dots, \\
 & x_3=\frac{K \tau}{6}s^3+\frac{1}{24}(2K_s\tau+K\tau_s)s^4+\dots \ .
\end{split}
\end{equation}
\\
At ordinary point where $K$, $\tau$ and $K_x$, $\tau_x$ are bounded the first terms in r.h.s. of  (\ref{xxx}) dominate and the curve, locally, is a twisted cubic. Its projections on the osculating plane  (spanned by $\vec{t}$ and $\vec{n}$), normal plane  (spanned by $\vec{b}$ and $\vec{n}$) and rectifying plane (spanned by $\vec{t}$ and $\vec{b}$) are parabola ($x_2-\frac{K}{2}x_1^2=0$), cusp ($9Kx_3^2-2\tau^2 x_2^3=0$) and cubic ($x_3+\frac{K\tau}{6}x_1^3=0$) respectively (fig. \ref{fig-pcc}, see e.g. \cite{Eis}).
\par
Characteristic feature of a curve (and a filament) in an ordinary point (i.e., outside the points of gradient catastrophe) is that it lies always on one side of positive direction of the normal and on one side of rectifying plane. \par
\begin{figure}
\centering
\includegraphics[width=4.5cm, height=5.5cm]{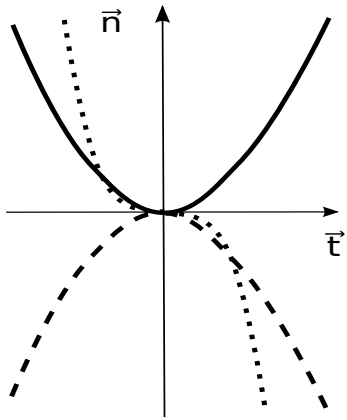}
\caption{Different projections on the osculating plane of a vortex filament near to the gradient catastrophe point.}
\label{fig-nt}
\end{figure}
\begin{figure}
\centering
\includegraphics[width=6cm, height=4.5cm]{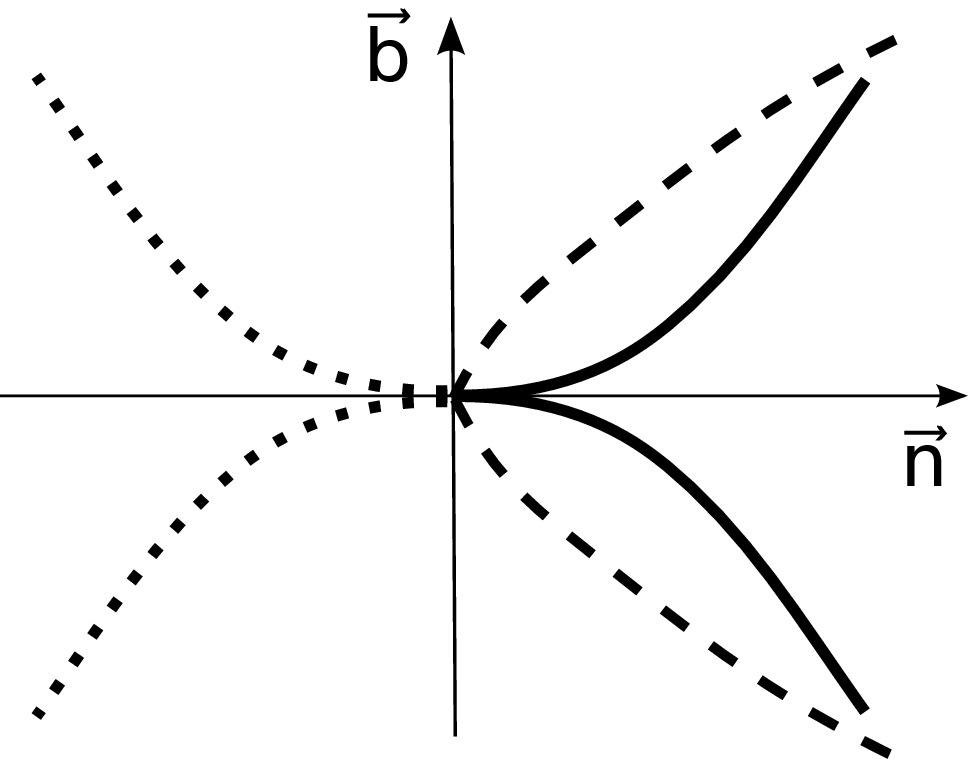}
\caption{Different projections on the normal plane of a vortex filament near to the gradient catastrophe point.}
\label{fig-bn}
\end{figure}

At the point $x_0$ of gradient catastrophe the behavior of a curve  changes drastically. Indeed, when $K_s$ and $\tau_s$ (i.e. $\epsilon K_x$ and $\epsilon \tau_y$) become large, the r.h.s. of the formula (\ref{xxx}) contains two parameters: small $s$ and large $K_s$ and $\tau_s$. So, the second terms in $x_2$ and $x_3$ become relevant and different balance between $s$ and $K_s$ ($\tau_s$) lead to different regimes in behavior of a curve around the origin. For instance, parabola in the osculating plane may
change sign or even convert into cubic curve  (fig. \ref{fig-bn}) while in the
normal plane it could be a plane $(3,4)$ curve and so on  (fig. \ref{fig-nt}). 
So, around the point $x_0$ of gradient catastrophe a filament oscillates from one side of the rectifying 
 plane to another and back. Such oscillation is quite similar to that of airfoil (see e.g. \cite{MilT}) and one can refer to these oscillations of a filament in the point of gradient catastrophe as a flutter. \par Another peculiarity of the point of gradient catastrophe can be seen in the behavior of the focal curve associated with filaments. The focal curve (see e.g. \cite{Eis}) is formed by the totality of centers of osculating spheres. Osculating sphere has a contact of third order with the curve in a points. Its center has coordinates
\begin{equation}
\label{X-osc}
 \vec{X}_o=\vec{X}+\frac{1}{K} \vec{n}-\frac{K_s}{K^2\tau}\vec{b}
\end{equation}
where $\vec{X}$ is the position vector of a point on the curve. The radius $R$ of the osculating sphere is given by
\begin{equation}
\label{R-osc} 
R^2=\frac{K^2\tau^2+K_s^2}{K^4\tau^2}.
\end{equation}
For a set of regular points the corresponding piece of the focal curve is regular too and it lies on a finite distance from the original curve.\par
\begin{figure}
\centering
\includegraphics[width=10cm, height=8cm]{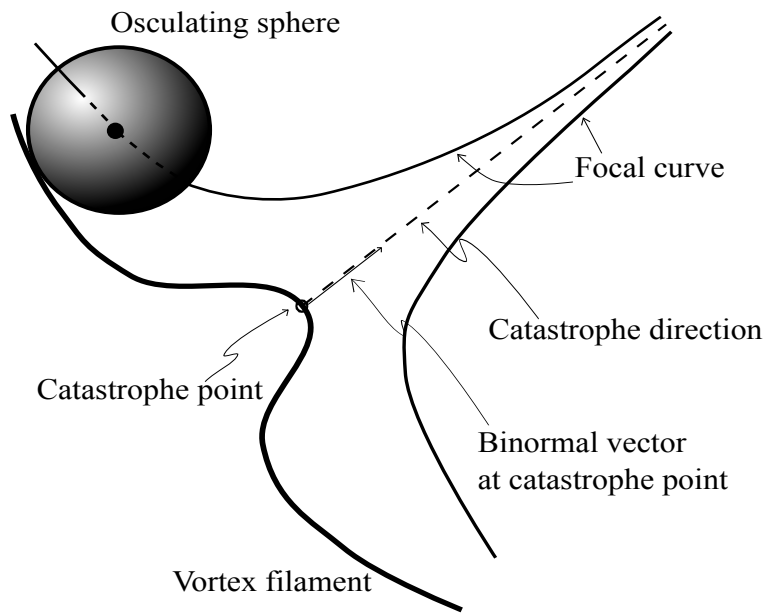}
\caption{The behavior of the focal curve near to the catastrophe point.}
\label{fig-cat-dir}
\end{figure}

At the point of gradient catastrophe $K_s \to \infty$ and, hence, the center of the osculating sphere as well as its radius blow up to infinity in the direction of the binormal vector (catastrophe direction, see figure \ref{fig-cat-dir}).
So, the point of gradient catastrophe for the original curve at which both curvature and torsion are finite corresponds to a point of the focal curve at which the second focal curvature ($-K_s/K^2\tau$)  becomes singular. We emphasize that even if our consideration for the original curve is purely local at the point of gradient catastrophe one observes a nonlocal effect of unbounded increase of the distance between a point on the original curve and the corresponding point on the associated focal curve.\par
 We note that the blow-up of $\vec{X}_0$ and $R$ for the focal curve happens also at the flattening points at which $\tau=0$ (for a discussion on the flattening points see e.g. \cite{Uri}). It would be of interest to understand fully a difference between the flattening points on a curve and points with unbounded $K_s$ in pure geometrical terms. As an obvious difference one may observe that at the flattening point the projection of the curve on the osculating plane remains to be a parabola and the velocity $\vec{V}$ and acceleration $\vec{V}_t$ remain bounded too.
\section{Elliptic umbilic and higher catastrophes}
In order to understand better the phenomenon described above one should analyze in more detail the structure of the singularity and the regularizing mechanism for the gradient catastrophe. For the focusing NLS equation such an analysis based on the $\epsilon$-expansion of the integrals of motions of the NLS/Toda equation has been performed in \cite{DGK}. Here we will follow a different approach proposed recently in \cite{KMA}.\par
Thus, we consider a neighborhood of the gradient catastrophe point $x_0,y_0$ and denote the values of $\beta$ at this point by $\beta_0$. Following the double scaling limit method (see e.g. \cite{DGZJ}-\cite{MAM}) we will look for solutions of the  system (\ref{KT-sys}) near the point of gradient catastrophe  in the form
\begin{equation}
 \label{gc-point}
\begin{split}
x=x_0+\varepsilon^{\alpha} x^*, \\
y=y_0+\varepsilon^{\sigma} y^*, \\
\beta=\beta_0+\varepsilon^{\eta} \beta^* \\ 
\end{split}
\end{equation}
where $\varepsilon \ll 1$ and numbers $\alpha,\sigma,\eta$ should be fixed by further consideration. 
In our approach the function $W$ is the principal object for such an analysis. For the system (\ref{KT-sys}) 
\begin{equation}
 \label{Wg}
W=W_1 x +W_2 y + \wW (\beta,\obeta)
\end{equation}
where
\begin{equation}
 \begin{split}
  W_1 &=\frac{1}{2}(\beta+\obeta), \\
 W_2 &=\frac{1}{8}(3\beta^2+2\beta\obeta+3\obeta^2)
+\frac{\gamma_1}{16}(5\beta^3+3\beta^2\obeta+3\obeta^2\beta+5\obeta^3)+
\gamma_2 \ln\left(\frac{\beta-\obeta}{2i}\right).
 \end{split}
\end{equation}
For the solutions of the system (\ref{KT-sys})  one has
\begin{equation}
 W_\beta=W_\obeta=0
\end{equation}
and also, for $K \neq 0$, equation (\ref{WBBb}), i.e.  $W_{\beta\obeta}=0$. Furthermore differentiating EPD equation (\ref{EPD-W}) with respect to $\beta$, one has 
\begin{equation}
 2 W_{\beta \obeta} +2(\beta-\obeta)W_{\beta\beta\obeta}= W_{\beta}-W_{\obeta}.
\end{equation}
Hence, at the point of gradient catastrophe $\beta_0$, $\obeta_0$ due to equation (\ref{WBBb}) and $W^0_{\beta \beta}=0$, one gets
\begin{equation}
 W_{\beta\beta\obeta}^0= W_{\beta\obeta\obeta}^0= 0
\end{equation}
where $f^0:=f(\beta_0,\obeta_0)$. So, at the expansion (\ref{gc-point}) for the function $W$, the third order derivatives $W_{\beta\beta\beta}^0$ and $W_{\obeta\obeta\obeta}^0$ are the only nonvanishing third order derivatives in the singular sector $\mathcal{M}^{sing}_1$.\par
As a result, an expansion of the function $W$ near the point of gradient catastrophe is given by
\begin{equation}
\label{WW0}
 \begin{split}
  W=&W_0+ \varepsilon^\alpha W_1^0 x^*+ \varepsilon^\sigma W_2^0 y^* +\frac{1}{2}\varepsilon^{\alpha+\eta} x^* (\beta^*+\obeta^*) +
+ \varepsilon^{\eta+\sigma} \left(({W_2}_{\beta}^0+{\wW}_{\beta}^0)\beta^* + c.c.\right) y^*   + \\&
 + \frac{\varepsilon^{2\eta+\sigma}}{2} \left[ ({W_2}_{\beta \beta}^0 +{\wW}_{\beta \beta}^0){\beta^*}^2 +
 ({W_2}_{\beta \obeta}^0 +{\wW}_{\beta \obeta}^0)\beta^*\obeta^* + c.c.\right] y^*
 +\frac{\varepsilon^{3\eta}}{6} \left({W}_{\beta \beta \beta}^0 {\beta^*}^3 +c.c.\right) + \dots \ .
\end{split}
\end{equation}   
Consequently
\begin{equation}
\label{Wbstar} 
\begin{split}
  W_{\beta^*}=&\frac{1}{2}\varepsilon^{\alpha+\eta} x^* 
+ \varepsilon^{\eta+\sigma} ({W_2}_{\beta}^0+{\wW}_{\beta}^0) y^*   
 + \frac{\varepsilon^{2\eta+\sigma}}{2} \left[ 2({W_2}_{\beta \beta}^0 +{\wW}_{\beta \beta}^0){\beta^*} +
 ({W_2}_{\beta \obeta}^0 +{\wW}_{\beta \obeta}^0)\obeta^* \right] y^* \\&
 +\frac{\varepsilon^{3\eta}}{2} {W}_{\beta \beta \beta}^0 {\beta^*}^2 + \dots =0  \ .
\end{split}
\end{equation}
For generic point in $\mathcal{M}^{sing}_1$ the third term is smaller than the second one and the nontrivial balance of three terms in (\ref{Wbstar}) is achieved if $\alpha+\eta=\eta+\sigma=3\eta$, i.e. $\alpha=\sigma=2\eta$.\par
Thus, near to the point of gradient catastrophe 
\begin{equation}
\label{W3}
 W=W_0+\varepsilon^{2\eta} \left( W_1^0 x^* + W_2^0 y^* \right) + \varepsilon^{3\eta} W^*
\end{equation}
where 
\begin{equation}
\label{Wstar}
W^*(\beta,\beta^*)=z^* \beta^* +\overline{z}^* \obeta^* + \frac{1}{6} a {\beta^*}^3+ \frac{1}{6} \overline{a} {\obeta^*}^3
\end{equation} 
and 
\begin{equation}
\label{zstar}
\begin{split}
 z^*=\frac{1}{2}x^* + ({W_2}^0_\beta +{\wW}^0_\beta ) y^*, \qquad
 a=W_{\beta\beta\beta}^0.
\end{split}
\end{equation}
Denoting $a^{1/3}\beta^* = U+iV$, one has
\begin{equation}
\label{EllUmbcat}
 W^*=fU+gV+\frac{1}{3}U^3-UV^2
\end{equation}
where $f$ and $g$ are certain linear combinations of $x^*$ and $y^*$.  In Thom's catastrophe theory language a function of this form is known as describing elliptic umbilic catastrophe \cite{Thom}. \par
The type of behavior of $\beta_x$ and $\beta_y$ near to the point of gradient catastrophe follows from (\ref{bxby}) and (\ref{Wbstar}). Indeed, near to the point $x_0,y_0$, one has 
\begin{equation}
 0=W_{\beta}=\varepsilon^{3\eta} W^*_{\beta}=\varepsilon^{2\eta} W^*_{\beta^*}=z^*+\frac{1}{2} a {\beta^*}^2.
\end{equation}
Hence $\beta^* = \left(-\frac{2 z^*}{a} \right)^{1/2}$ and
\begin{equation}
\label{bz-cat}
 \beta_z = \varepsilon^{-\eta} \beta^*_{z^*}=  - \varepsilon^{-\eta} \left(-a z^*/2 \right)^{-1/2}
 \end{equation}
\par The elliptic umbilic singularity (\ref{EllUmbcat}) and formula (\ref{bz-cat}) describe generic point of gradient catastrophe which corresponds to the sector $\mathcal{M}^{sing}_1$. For variable parameters $\gamma_{1}$, $\gamma_{2}$ and other parameters describing the family of initial data, higher singularities corresponding to the singular sectors $\mathcal{M}^{sing}_n$ discussed in section 5 may appear.\par
For instance, if for some $\gamma_{1}$, $\gamma_{2}$, $W_{\beta\beta\beta}^0=0$ and $W_{\beta\beta\beta\beta}^0 \neq 0$ then instead of the expansions (\ref{WW0}) one has
\begin{equation}
\label{WW0-4}
 \begin{split}
  W=&W_0+ \varepsilon^\alpha W_1^0 x^*+ \varepsilon^\sigma W_2^0 y^* +\frac{1}{2}\varepsilon^{\alpha+\eta} x^* (\beta^*+\obeta^*) 
+ \varepsilon^{\eta+\sigma} \left(({W_2}_{\beta}^0+{\wW}_{\beta}^0)\beta^* + c.c.\right) y^*   + \\&
 + \frac{\varepsilon^{2\eta+\sigma}}{2} \left[ ({W_2}_{\beta \beta}^0 +{\wW}_{\beta \beta}^0){\beta^*}^2 +
 ({W_2}_{\beta \obeta}^0 +{\wW}_{\beta \obeta}^0)\beta^*\obeta^* + c.c.\right] y^*
 +\frac{\varepsilon^{4\eta}}{24} \left({W}_{\beta \beta \beta \beta}^0 {\beta^*}^4 +c.c.\right) + \dots \ .
\end{split}
\end{equation}   
So,
\begin{equation}
\label{Wbstar-4} 
\begin{split}
  W_{\beta^*}=&\frac{1}{2}\varepsilon^{\alpha+\eta} x^* 
+ \varepsilon^{\eta+\sigma} ({W_2}_{\beta}^0+{\wW}_{\beta}^0) y^*   
 + \frac{\varepsilon^{2\eta+\sigma}}{2} \left[ 2({W_2}_{\beta \beta}^0 +{\wW}_{\beta \beta}^0){\beta^*} +
 ({W_2}_{\beta \obeta}^0 +{\wW}_{\beta \obeta}^0)\obeta^* \right] y^* \\&
 +\frac{\varepsilon^{4\eta}}{6} {W}_{\beta \beta \beta \beta}^0 {\beta^*}^3 + \dots =0  \ .
\end{split}
\end{equation}
and, hence, in this case
\begin{equation}
 \alpha=\sigma=3\eta.
\end{equation}
Thus, at the sector $\mathcal{M}^{sing}_2$ one has
\begin{equation}
 W=W_0+\varepsilon^{3\eta} \left( W_1^0 x^* + W_2^0 y^* \right)+\varepsilon^{4\eta} W^*+\dots
\end{equation}
where
\begin{equation}
 W^*=z^* \beta^* + \overline{z}^* \obeta^* +\frac{1}{24} \tilde{a} {\beta^*}^4+\frac{1}{24} \overline{\tilde{a}} {\obeta^*}^4
\end{equation}
and $z^*=\frac{1}{2} x^*  +({W_2}_{\beta}^0+{\wW}_{\beta}^0) y^* $  as in the previous case (when ${W}_{\beta \beta \beta}^0 \neq 0$), 
while now  $\tilde{a}={W}_{\beta \beta \beta \beta}^0$. \par
Denoting $\tilde{a}^{1/4} \beta^* =\widetilde{U} + i \widetilde{V} $, one gets
\begin{equation}
W^*=l \widetilde{U}+m \widetilde{V} + \frac{1}{12}\widetilde{U}^4- \frac{1}{2}\widetilde{U}^2\widetilde{V}^2+ \frac{1}{12}\widetilde{V}^4.
\end{equation}
The last three terms reproduce the canonical form of a (nonsimple) unimodular parabolic singularity $X_9$ in terms of Arnold's classification.\par 
Near to the point of gradient catastrophe $x_0,y_0$ the behavior of $\beta^*$ is given by
\begin{equation}
 0=W_\beta=W^*_{\beta^*}= {z}^*  +\frac{1}{6} a {\beta^*}^3.
\end{equation}
 Therefore $\beta^* = \left(-\frac{6 z^*}{a}\right)^{1/3}$ and
\begin{equation}
 \beta^*_{z^*} = \frac{1}{3}\left(-\frac{6 }{a{z^*}^2}\right)^{1/3}.
\end{equation}
In the sector $\mathcal{M}^{sing}_n$ generically one has $\alpha=\sigma=(n+2)\eta$ and therefore
\begin{equation}
 W=W_0+\varepsilon^{(n+2) \eta} W^*
\end{equation}
where 
\begin{equation}
\label{Wstargen}
 W^*=z^* \beta^* + \overline{z}^* \obeta^* +\frac{1}{(n+2)!} a_n {\beta^*}^{n+2}+\frac{1}{(n+2)!} \overline{a_n} {\obeta^*}^{n+2}
\end{equation}
with $a_n = (\partial_\beta^{n+2} W)^0$ and
\begin{equation}
 \beta_{z^*} \sim {z^*}^{-n/(n+1)}.
\end{equation}
Appearance of these singularities and particularly the time order of their appearance strongly depends on concrete structure and properties of  vortex filament.\par
For nongeneric point of singular sector one might have a different behavior of function $W$. For instance, for the singular sector  $\mathcal{M}^{sing}_1$ in situation when
\begin{equation}
\label{xy-diff}
 {W_2}_{\beta}(\beta_0,\obeta_0) + \wW_\beta(\beta_0,\obeta_0)=0,
\end{equation}
the third term in (\ref{WW0})  of the order $2\eta+\sigma$ becomes relevant and the balance of order requires that $\alpha+\eta=\sigma+2\eta=3\eta$. Hence, in such situation $\alpha=2\eta$, $\sigma=\eta$ and instead of the expression (\ref{Wbstar}) one has
\begin{equation}
\label{WW0-new}
 \begin{split}
  W=&W_0+ \varepsilon^\eta W_2^0    y^*  + \varepsilon^{2\eta} \frac{1}{2}(\beta_0+\obeta_0) x^*   +\varepsilon^{3\eta} W^*.
\end{split}
\end{equation}
where
\begin{equation}
\label{Wstar-new}
 \begin{split}
  W^*=& \frac{1}{2}x^* (\beta^*+\obeta^*) 
 + \frac{1}{2} \left[ ({W_2}_{\beta \beta}^0 +{\wW}_{\beta \beta}^0){\beta^*}^2 +
 ({W_2}_{\beta \obeta}^0 +{\wW}_{\beta \obeta}^0)\beta^*\obeta^* + c.c.\right] y^* \\&
 +\frac{\varepsilon^{3\eta}}{6} \left({W}_{\beta \beta \beta}^0 {\beta^*}^3 +c.c.\right) + \dots \ .
\end{split}
\end{equation}
So, in the case (\ref{xy-diff}) the variables $x^*$ and $y^*$ play a quite different role in contrast to the generic case  (\ref{Wstar}), (\ref{zstar}). Such type of behavior will be studied elsewhere.
\section{Double scaling regularization. Painlev\'e I and other equations}
At the point of gradient catastrophe the quasiclassical approximation ceases to be valid and contribution of the higher order derivatives terms in the governing equation becomes relevant. A standard way to deal with this problem is to begin with the full (dispersionfull) version of the governing equation (i.e. Da Rios system (\ref{bin-intrinsic}) or intrinsic version of equation (\ref{Hirota})) and, then, performing an appropriate double scaling limit, calculate the required regularization terms. For the NLS equation (\ref{NLS}) such an analysis has been performed in \cite{DGK}. \par Here we will follow another approach discussed in \cite{KMA,KOflutter}. The basic idea is to incorporate higher order derivatives terms without any reference to the exact (dispersionfull) equation. \par The first step is based on the observation (section 3)  that governing equations describe critical points of the function $W$. Thus, in order to incorporate derivatives of $\beta$, it is quite natural to modify $W$ to $W_q(\beta,\obeta,\beta_x,\obeta_x,\beta_y,\obeta_y)$ in such a way that equation (\ref{crit-eqn}) is substituted by the Euler-Lagrange equations
\begin{equation}
  \frac{\partial W_q}{\partial \beta}=0, \qquad \frac{\partial W_q}{\partial \obeta}=0   
\end{equation}
 for the functional $\int W_q dzd\overline{z}$.\par
Second point is to use the specific properties of the function $W$. Namely, the real-valuedness of $W$ 
\begin{equation}
 \overline{W(\beta,\obeta)}=W(\beta,\obeta)
\end{equation}
  and the fact that in the perturbations $W^*$ (see formulas (\ref{Wstar}), (\ref{Wstargen})) the contributions of $\beta^*$ and $\obeta^*$ are separated. It is natural to look for the modifications of $W$ which preserve these properties. Thus, near the point of gradient catastrophe one has the following $W_q$ (omitting inessential terms in (\ref{W3}) and (\ref{Wstar}) and choosing $\eta=1$) 
\begin{equation}
 W_q=  W_0+\epsilon^3\left[ \frac{1}{2}(z^* \beta^*+\overline{z}^*\overline{\beta}^*)+\frac{1}{6}(a {\beta^*}^3+ \overline{a}{\overline{\beta}^*}^3) \right] + \frac{1}{2}\epsilon^\delta 
\left[
b {\beta_{z^*}^*}^2+ \overline{b}\, {\overline{\beta}_{\overline{z}^*}^*}^2
\right]
\end{equation}   
where $\delta$ and $b$ are appropriate constants. Euler-Lagrange equations
\begin{equation}
\begin{split}
 \frac{\partial W_q}{\partial \beta^*}  -\left( \frac{\partial W_q}{\partial \beta_{z^*}^*} \right)_{z^*}=0, \\
\frac{\partial W_q}{\partial \obeta^*}  -\left( \frac{\partial W_q}{\partial \obeta_{\overline{z}^*}^*} \right)_{\overline{z}^*}=0
\end{split}
\end{equation}
implies that $\delta=3$. So, one has the equation
\begin{equation}
\label{P-Iour}
 \frac{1}{2}\left[z^*+a {\beta^*}^2 \right] - 
b {\beta_{z^* z^*}^*}
=0.
\end{equation}
This equation is converted into the classical Painlev\'e I equation
\begin{equation}
\label{P-Ibis}
\Omega_{\xi\xi}=6\Omega^2-\xi.
\end{equation}
by the change of variables 
\begin{equation}
 z^*=\lambda \xi, \qquad \beta^*=-\frac{\lambda^3}{2b}\Omega
\end{equation}
with $a\lambda^5=-24b^2$.\par
So, the regularized behavior at the point of the gradient catastrophe for the vortex filament dynamics is governed by the Painlev\'e I equation. The relevance of this equation for the NLS/Toda system near to the critical point has been first observed in a different way in \cite{DGK}. The correspondence between the solution of equation (\ref{P-Iour}) and those given by the formula (5.21) in \cite{DGK} is (at $b=1$) $u=K^2$, $v=2\tau$ and
\begin{equation}
 \begin{split}
  \overline{x}=&-i \epsilon^{4/5} \left( r e^{i \psi}\right)^{-1/5} \left(-\frac{a}{36 K_0} \right)^{1/5}z^*,\\
  \overline{\beta}=&-ib \epsilon^{2/5} \left( r e^{i \psi}\right)^{2} \left(-\frac{a}{36 K_0} \right)^{3/5}\beta^*
 \end{split}
\end{equation}
where $\overline{x}$, $\overline{\beta}$, $r$ and $\psi$ are defined in (2.19) and (2.22) of \cite{DGK}.
One also concludes that $\epsilon=\varepsilon^{2/5}$. \par
In the paper (\cite{DGK}) it was conjectured that any generic solution of the NLS equation near to the critical point is described by the tritronque\'e solution of the Painlev\'e I equation. Extension of this observation to our case allows us to formulate the 
\begin{conj}
Generic behavior of the vortex filament near the point $x_0,y_0$ of gradient catastrophe within the localized induction approximation
 is given by
\begin{equation}
\label{b-conj}
 \beta(s,t_0,\xi) \simeq \beta_0 
-\varepsilon \left(-\frac{36}{a^3b} \right)^{1/5} \Omega_0\left(\left(-\frac{a}{12 b^3} \right)^{1/5} \frac{s-s_0}{\varepsilon^3} \right)
\end{equation}
where $\Omega_0$ is the tritronque\'e solution of the Painlev\'e I equation (\ref{P-Ibis}). 
\end{conj}
Such an approach can be extended also to the singular sectors $\mathcal{M}^{sing}_n$. In these sectors the modified $W_q$, for example,
 can be chosen in the following form
\begin{equation}
 W_{qn}=W_0+\varepsilon^n \left( \frac{1}{2}(z^* \beta^*+\overline{z}^*\overline{\beta}^*) +\frac{1}{(n+2)!}a_n {\beta^*}^{n+2}+\frac{1}{(n+2)!}\overline{a_n} {\obeta^*}^{n+2} \right)+\varepsilon^{\delta_n} \left( \frac{1}{2}b_n {\beta^*}_{z^*}^{2}+\frac{1}{2}\overline{b_n}\ {\obeta^*}^{2}_{\overline{z}^*} \right) 
\end{equation} 
where $\delta_n$ are appropriate constants. The Euler-Lagrange equation readily implies that $\delta_n=n$. So, the modified $W$ is given by
\begin{equation}
 W_{qn}=W_0+\varepsilon^n \left(\frac{1}{2}(z^* \beta^*+\overline{z}^*\overline{\beta}^*) +\frac{1}{(n+2)!}a_n {\beta^*}^{n+2}+\frac{1}{(n+2)!}\overline{a_n} {\obeta^*}^{n+2} + \frac{1}{2}b_n {\beta^*}_{z^*}^{2}+\frac{1}{2}\overline{b_n}\ {\obeta^*}^{2}_{\overline{z}^*}  \right) 
\end{equation} 
and the corresponding Euler-Lagrange equation for $\beta^*$ is
\begin{equation}
 b_n \beta_{z^*z^*}=\frac{a_n}{(n+1)!} {\beta^*}^{n+1}+\frac{z^*}{2}.
\end{equation}
By the appropriate change of variables 
\begin{equation}
 z^*=\lambda \xi, \qquad \beta^*=-\frac{\lambda^3}{2b_n}\Omega, \qquad 2a_n\lambda^{3n+2}=-(-2b_n)^{n+1}{(n+1)!}{(n+2)!}
\end{equation}
one gets the equation 
\begin{equation}
 \Omega_{\xi \xi} =(n+2)!\Omega^{n+1}-\xi.
\end{equation}
Eventual relation of these equations with appropriate limits of NLS hierarchy, higher Painlev\'e equations and other types of regularizations containing higher order derivatives will be considered in a separate paper.
\par
{\bf Acknowledgements} The authors are grateful to M. Bertola, B. Dubrovin and E. Ferapontov for useful discussions. This work has been partially supported by PRIN grant no 28002K9KXZ and by FAR 2009 (\emph{Sistemi dinamici Integrabili e Interazioni fra campi e particelle}) of the University of Milano Bicocca.

\end{document}